\documentclass{aa}
\usepackage{psfig,times}
 
\begin{document} \thesaurus{09 
Interstellar medium (ISM), nebulae
(09.09.1; 
 09.10.1; 
 09.19.2; 
 13.25.4)} 
   \title{X-ray observations of supernova remnant G54.1+0.3:
X-ray spectrum and the discovery of an X-ray jet}

   \author{F.J. Lu
          \inst{1,2}
          \and
          B. Aschenbach\inst{1}
	\and L.M. Song\inst{2}}

   \offprints{F.J. Lu (lufj@astrosv1.ihep.ac.cn)}

   \institute{Max-Planck-Institut f\"ur Extraterrestrische Physik, 
D-85740 Garching bei M\"unchen, Germany
         \and
             Laboratory of Cosmic Ray and High Energy Astrophysics, 
Institute of High Energy Physics, CAS, Beijing 100039, China\\}

   \date{Received , ; accepted ,}

   \maketitle \markboth{Lu, Aschenbach $\&$ Song: supernova remnant G54.1+0.3}{}

   \begin{abstract}
We present in this paper analyses of the $ROSAT$ PSPC and $ASCA$ SIS and GIS observations
 of the Crab-like supernova remnant (SNR) G54.1+0.3. Its spectrum 
 obtained by $ROSAT$ PSPC favors a
power law model with  a photon index of 
-0.8$^{+0.8}_{-2.0}$, absorbed energy flux in 0.1-2.4 keV of 
1.0$\times 10^{-12}$ erg cm$^{-2}$ s$^{-1}$, and absorption column density
of 12.3$^{+8.0}_{-3.2}\times 10^{21}$ cm$^{-2}$. $ASCA$ SIS observation
shows that its spectrum can also be best fitted with power law model. The
fitted parameters are, photon index -1.9$^{+0.1}_{-0.2}$, absorbed energy
flux in 0.7-2.1 keV 6.5$\times$10$^{-13}$ erg cm$^{-2}$ s$^{-1}$, and
column density 17.9$^{+2.8}_{-2.5}\times 10^{21}$ cm$^{-2}$. 
The high absorption column
density indicates a distance similar to the radius of the galaxy.
The 0.1-2.4 keV X-ray luminosity of G54.1+0.3 
is 3.2$\times$10$^{33}$$d_{10}^2$ erg  s$^{-1}$, where $d_{10}$ is the 
distance in 10 kpc. 
With an image restoration method we
have obtained high spatial resolution X-ray image of the remnant, which
clearly shows an X-ray jet pointing to the northeast with a length
about 40$\arcsec$ from the center of the nebula. Its X-ray luminosity
in 0.1-2.4 keV is about 5.1$\times$10$^{32}$$d_{10}^2$ erg s$^{-1}$. The X-ray 
jet is consistent with the radio extension
to the northeast in both direction and position.  
We propose that the X-ray jet is connected with 
the pulsar assumed to exist in the remnant.   

      \keywords{X-ray: ISM -- ISM: supernova remnants -- ISM:jet and outflows
--ISM: individual: G54.1+0.3 }
 \end{abstract}

%

\section{Introduction}

Radio source G54.1+0.3 was first suggested to be a Crab-like SNR by 
Reich et al. (1985) for its flat spectral index of $\alpha\sim-0.1\pm0.1$,
filled-center morphology and significant polarization. This identification
to G54.1+0.3 was confirmed by Velusamy $\&$ Becker (1988) with high resolution 
multifrequency observations with the VLA and OSRT. In the high resolution
VLA maps, G54.1+0.3 has a filled-center brightness distribution peaks around
R.A.(2000) =19:30:30, 
DEC(2000)=18:52:11 and extends to the 
northeast and north (Velusamy \& Becker 1988). They pointed out that these
 extensions
 are reminiscent of the 
radio jets seen in the Crab (Velusamy 1984), CTB80 (Angerhofer et al. 1981) 
and G332.4+0.1 (Roger et al. 1985). 

X-rays from G54.1+0.3 was detected by $EINSTEIN$ 
IPC (resolution $\sim$ 1$\arcmin$) with a source strength
of 0.016$\pm$0.004 counts s$^{-1}$ in the energy band 0.5-4.0 keV (Seward 1989). 
No extent to the X-ray emission was found, due to both its small
angular size (2.0$^{\prime}\times1.2^{\prime}$) (Velusamy \& 
Becker 1988) and its low flux. A power law spectral fitting with 
energy index of 1.0 gives column density N$_H$ between 5$\times$10$^{21}$
and 1$\times$10$^{23}$ cm$^{-2}$, with the best fit value 
of 3$\times$10$^{22}$ cm$^{-2}$, indicating a large distance of this source.   

In the paper we present the analyses of $ROSAT$ PSPC 
and $ASCA$ GIS and SIS observations of 
G54.1+0.3. We obtain its spectral information, and, with the aid of an 
image restoration method, we obtain a high spatial resolution X-ray map 
of the remnant which clearly shows an X-ray jet pointing to the northeast.

\section{Observations and analysis method}
The $ROSAT$ PSPC pointing observation of SNR G54.1+0.3 was carried out from
April 11th to 18th, 1991 with a total acceptable  observational time of 
20271 seconds. We use $EXSAS$ (Zimmermann et al. 1998) to analyze its
spectrum and produce a 0.1-2.5 keV X-ray image (figure 1) whose spatial 
resolution is the intrinsic resolution of PSPC (40$\arcsec$).

G54.1+0.3 was also observed with $ASCA$ observatory (Tanaka et al. 1994)
 continuously from April 27th to 28th, 1997, using the two Gas Imaging
 Spectrometers (GIS-2 and GIS-3) and the two Solid State Imaging 
 Spectrometers (SIS-0 and SIS-1). Data were collected by the two
 GIS detectors with a photon time-of-arrival resolution of
 4.88$\times10^{-4}$ s in the high bit-rate modes. An effective exposure
 time 16.5 ks was achieved for each detector. The SIS detectors were
 operated in the 1-CCD faint mode in which read-out is every 4 s. All SIS 
 data were filtered using the standard screening criteria, which
 resulted in effective exposures of 19 ks and 20.7 ks for SIS-0 and
  SIS-1 respectively. Since the two GIS detectors were operated in the
  high time-of-arrival resolution model, we used the GIS data for
  temporal analysis. The SIS detectors which are sensitive to photons
  in 0.5-10.0 keV have superior energy resolution compared to the GIS, and
  so the SIS data are used for spectral analysis. 

Due to the small angular size (120$\arcsec\times75\arcsec$) of G54.1+0.3 
and the limited
spatial resolution ($\sim40\arcsec$) of PSPC, the PSPC observation can 
not directly give even a coarsely resolved image of G54.1+0.3.   
In order to obtain an image with higher spatial resolution, we use
the widely used Lucy-Richardson formula (Richardson 1972, Lucy 1974) 
to eliminate the point spread function effect in figure 1.
In the iteration process we have used the mean background 
as the lower limit constraints, in order to improve the quality of the 
restorated image, as done by Li \& Wu (1994), Lu et al. (1996) 
and Zhang et al. (1998).

\begin{figure}[thb]
 \psfig{figure=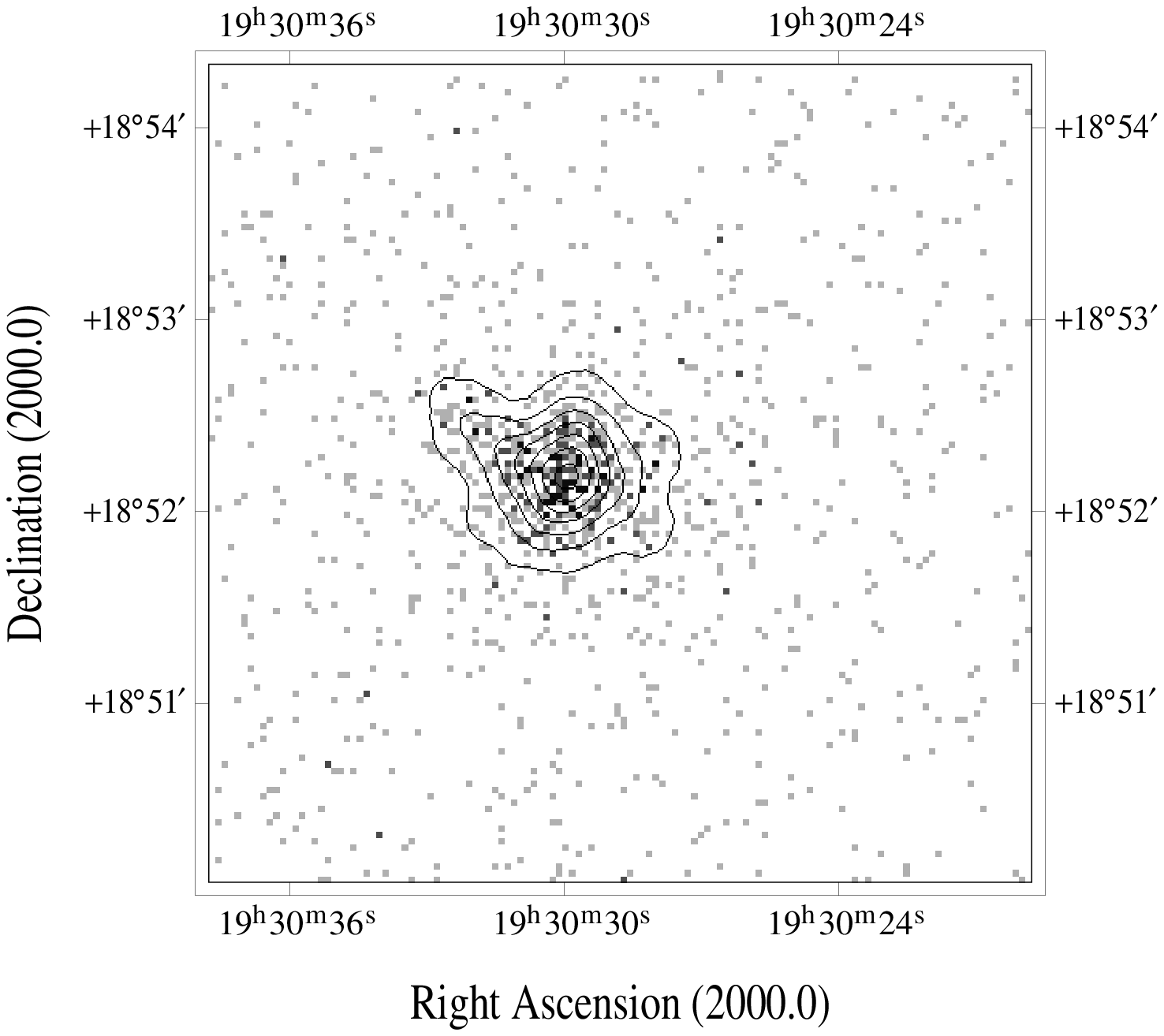,width=8.8truecm,angle=0,%
   bbllx=65pt,bblly=180pt,bburx=458pt,bbury=544pt,clip=}
\caption{Count-rate map in 0.1-2.5 keV obtained by $ROSAT$ PSPC 
observation to G54.1+0.3. North is up and east is left. The 
contours overlaid represent the smoothed count-rate map with a 10$\arcsec$
FWHM Gaussian filter. The contour intervals are linear in step 
of 10$^{-5}$ count s$^{-1}$ per 2$\arcsec\times$$2\arcsec$ pixel. The lowest
contour corresponds to a brightness level of 
10$^{-5}$ cnt s$^{-1}$ per pixel.}
\label{picture}
\vspace{0cm}
\end{figure}
\section{Results}
\subsection{Spectrum from $ROSAT$ PSPC observation}
The $ROSAT$ PSPC spectrum 
 of G54.1+0.3 shows a lack of low energy photons and
peaks at energy channel 150 (about 1.5 keV). The spectrum can be
fitted with power law model and Raymond-Smith (1977) thermal
plasma model. The power law model yields a photon index of 
-0.8 with 1 $\sigma$ error range of -2.8 to 0.0 and an absorption 
column density of 12.3$\times$10$^{21}$ cm$^{-2}$ with 1 $\sigma$ error
range of 8 to 20$\times$10$^{21}$ cm$^{-2}$ (see figure 2). The 
thermal plasma model derives a plasma temperature of 1.8 keV ($>$ 1.2keV)
and absorption column density of 21.1$\times$10$^{21}$ cm$^{-2}$ with
1 $\sigma$ error range of 15-26$\times$10$^{21}$ cm$^{-2}$. 
The reduced $\chi^2$ values are almost the same, 0.831 for power law model
and 0.834 for thermal plasma model. We adopt the power law
model in this paper for it gives the best and the most reasonable
fit to the $ASCA$ SIS spectrum, as shown in the next section.
It gives the absorbed and unabsorbed 
0.1-2.4 keV X-ray energy fluxes of 1.0$\times10^{-12}$ 
and 3.4$\times10^{-12}$ erg cm$^{-2}$ s$^{-1}$, respectively. 
Figure 3 shows the power law model fitting results.

\begin{figure}[thb]
 \psfig{figure=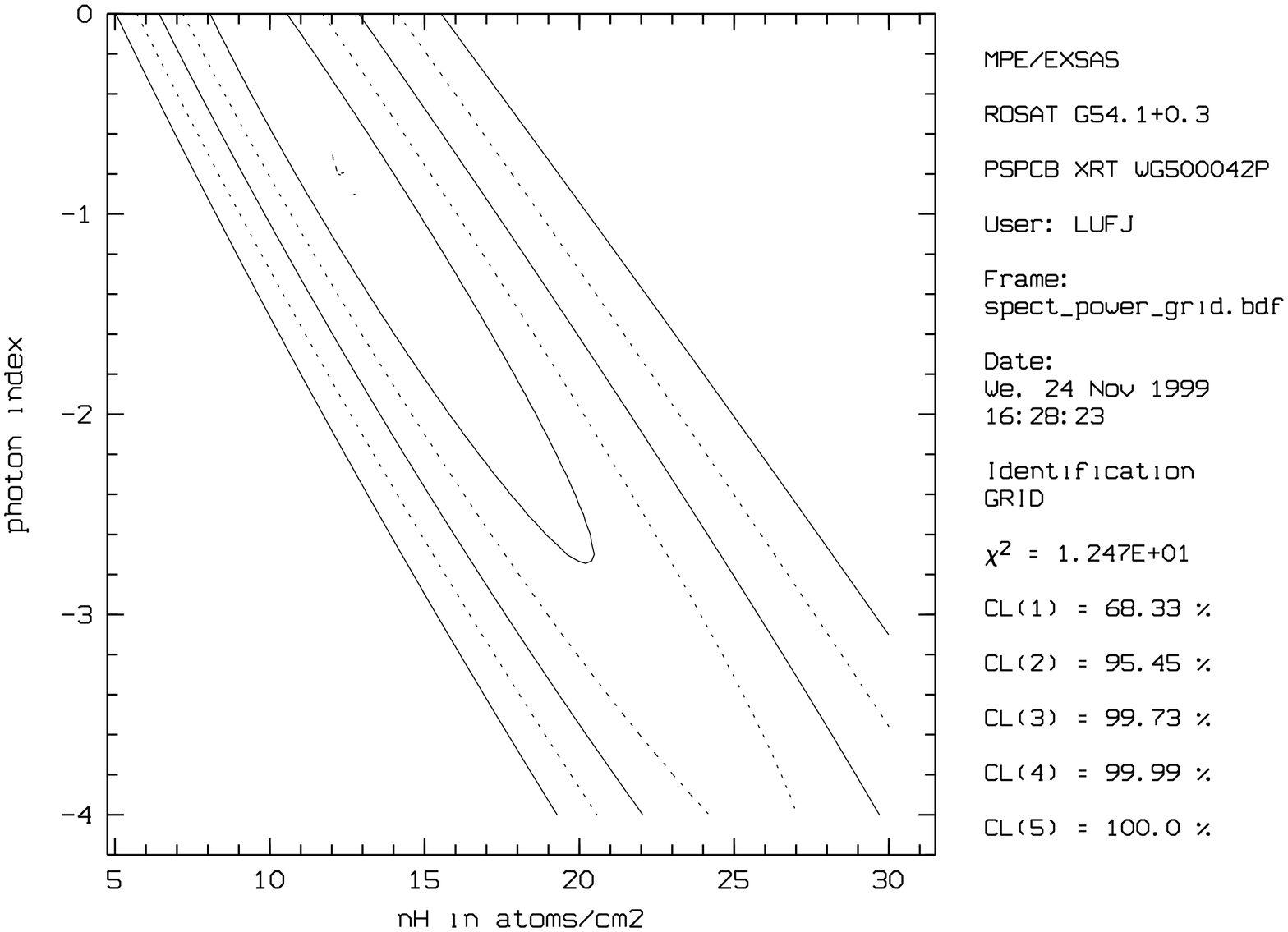,width=6.8truecm,height=5.truecm,angle=270,%
   bbllx=50pt,bblly=0pt,bburx=575pt,bbury=600pt,clip=}
\caption{$\chi^2$ distribution on the column density--photon index
plan of the power law spectral fitting of $ROSAT$ PSPC observation
of G54.1+0.3. Contours from the inner to the outer correspond to
1$\sigma$ to 5$\sigma$.} 
\label{picture}
\vspace{0cm}
\end{figure}

\begin{figure}[thb]
 \psfig{figure=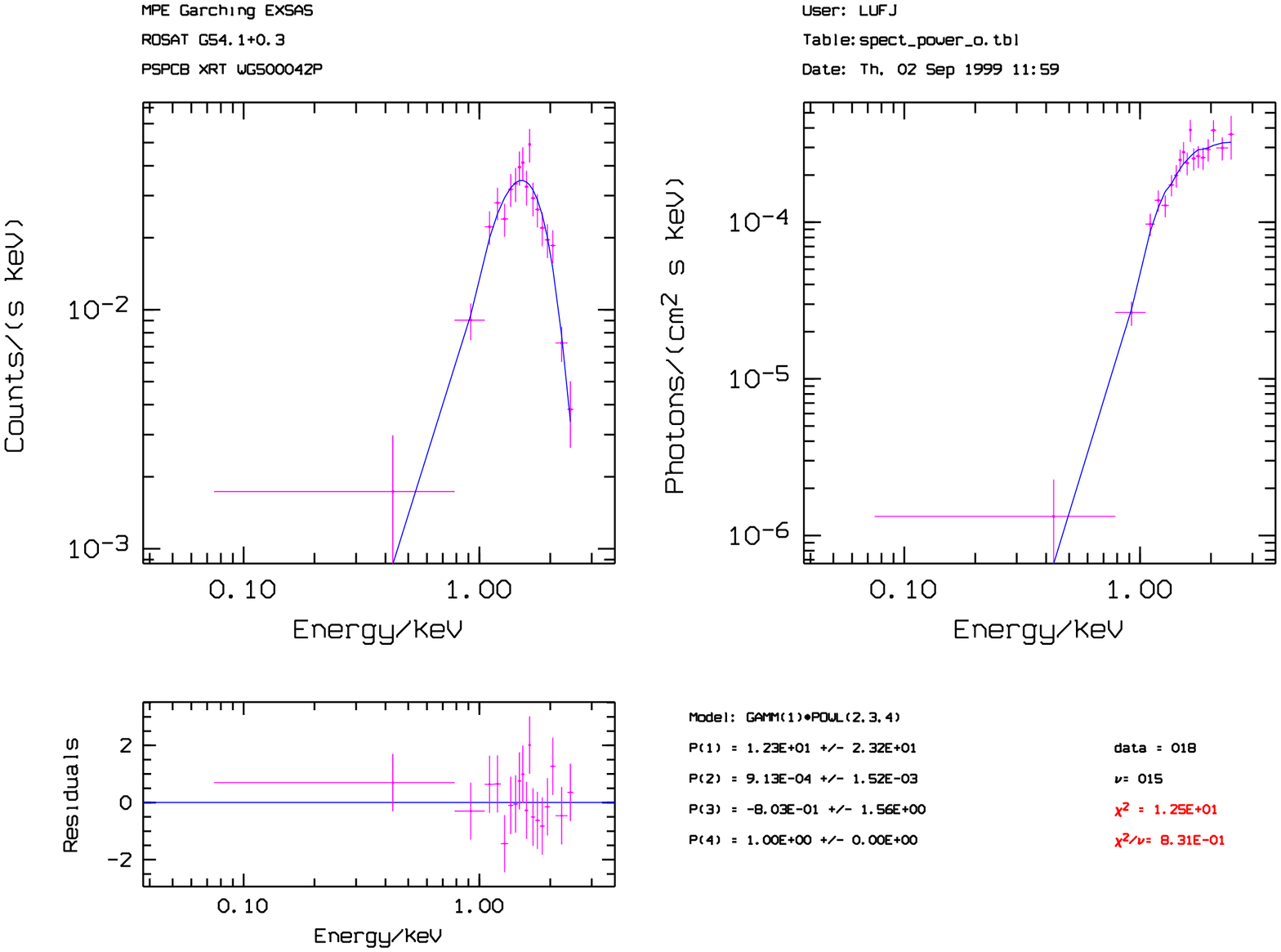,width=7.8truecm,height=7truecm,angle=270,%
   bbllx=85pt,bblly=30pt,bburx=550pt,bbury=400pt,clip=}
\caption{Spectral fitting results to the X-ray emission of G54.1+0.3
obtained by $ROSAT$ PSPC with a power law model. The parameters 
are presented in the text.}
\label{picture}
\vspace{0cm}
\end{figure}

\subsection{Spectrum from $ASCA$ SIS observation}
The SIS spectra of the source were extracted within a 4.5 arcminutes radius
region.
After subtracted the source
region, another region of the CCD was used for background subtraction.
The source and background spectra obtained from both SISs were added to
obtain improved statistics. The spectral anlyses software is XSPEC. 
Energies above 8 keV were not used because
of the poor signal to noise ratio. We have used power law, blackbody, 
single temperature bremsstrahlung and Raymond-Smith thermal plasma models
to fit the spectrum, and found that only the power law model and the 
thermal bremsstrahlung model give 
acceptable and reasonable fits. The obtained parameters 
of the power law model are: photon index $\alpha$ -1.9$^{+0.2}_{-0.2}$, column density
$N_H$ 17.9$^{+2.8}_{-2.5}\times10^{21}$ cm$^{-2}$, 0.7-2.1 keV energy
flux 6.5$\times$10$^{-13}$ erg cm$^{-2}$ s$^{-1}$, reduced $\chi^2$ 0.7. 
Parameters of a thermal bremstrahlung model are: temperature 
$T_e$ 7.9$^{+3.9}_{-3.1}$ keV, column density
$N_H$ 15.4$^{+2.0}_{-1.9}\times10^{21}$ cm$^{-2}$, 0.7-2.1 keV energy
flux 8.6$\times$10$^{-13}$ erg cm$^{-2}$ s$^{-1}$, reduced $\chi^2$ 0.8.
We choose the power law model in this paper for that it has the smallest $\chi^2$
and a power law X-ray spectrum is the typical property of the X-ray emission of
a Crab-like SNR.
For illustration,
we show in figure 4 the best fit power law model and its residuals.
\begin{figure}[thb]
 \psfig{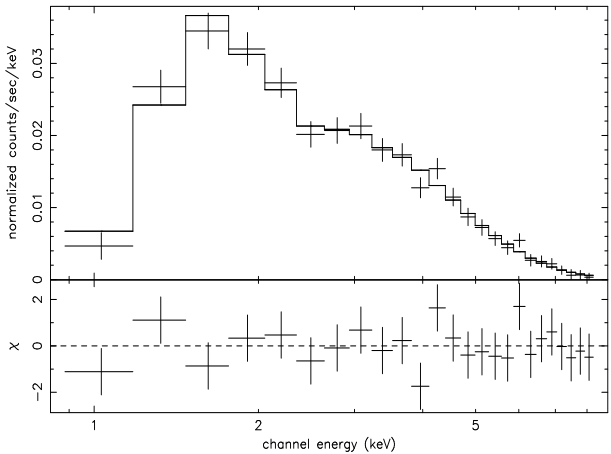}
\caption{Spectral fitting results to the X-ray emission of G54.1+0.3
obtained by $ASCA$ SIS with a power law model. 
The parameters are presented in the text.}
\label{picture}
\vspace{0cm}
\end{figure}

\subsection{Temporal analysis}
We examined the $ASCA$ GIS data for temporal variability by extracting
photons from a 6 arcminutes radius circle centered on the source.
A search for coherent pulsations from the source was made by combing the two
GIS high-time-resolution data sets (time resolution 4.88$\times$10$^{-4}$ s)
and the arrival times of the used 1805 photons were barycentered. 
We performed a 
restricted search for periodic signals between 0.01 s and 2 s using a 
folding technique (20 phase bins per fold), and detected no pulsation 
with a significance of more than 3$\sigma$ in this period range.

\subsection{Image restoration results}
The distribution of photons detected 
by $ROSAT$ PSPC peak at 1.5 keV and is quite symmetric. We thus use the PSPC
point spread function in 1.5 keV and the method described in section 2 to
restore the original image (figure 1). The iteration stops after 50
iterations (indeed the restored image is
insensitive to the iteration number after 20 iterations). The restored image
is shown in figure 5, in which a jet-like feature (hereafter JLF)  
pointing to the northeast appears, in addition to the $\sim$30$\arcsec$
diameter bright nebula coinciding 
with the brightest radio region. The angular distance from the head of
the JLF to the center of the central bright nebula is about 40$\arcsec$.
The total photon flux of the bright nebula 
is 2.07$\times$10$^{-2}$ counts s$^{-1}$, that of the JLF is 
about 3.9$\times$10$^{-3}$ counts s$^{-1}$, about 430 and 80 photons 
have been detected from the bright nebula and JLF respectively.
\begin{figure}[thb]
 \psfig{figure=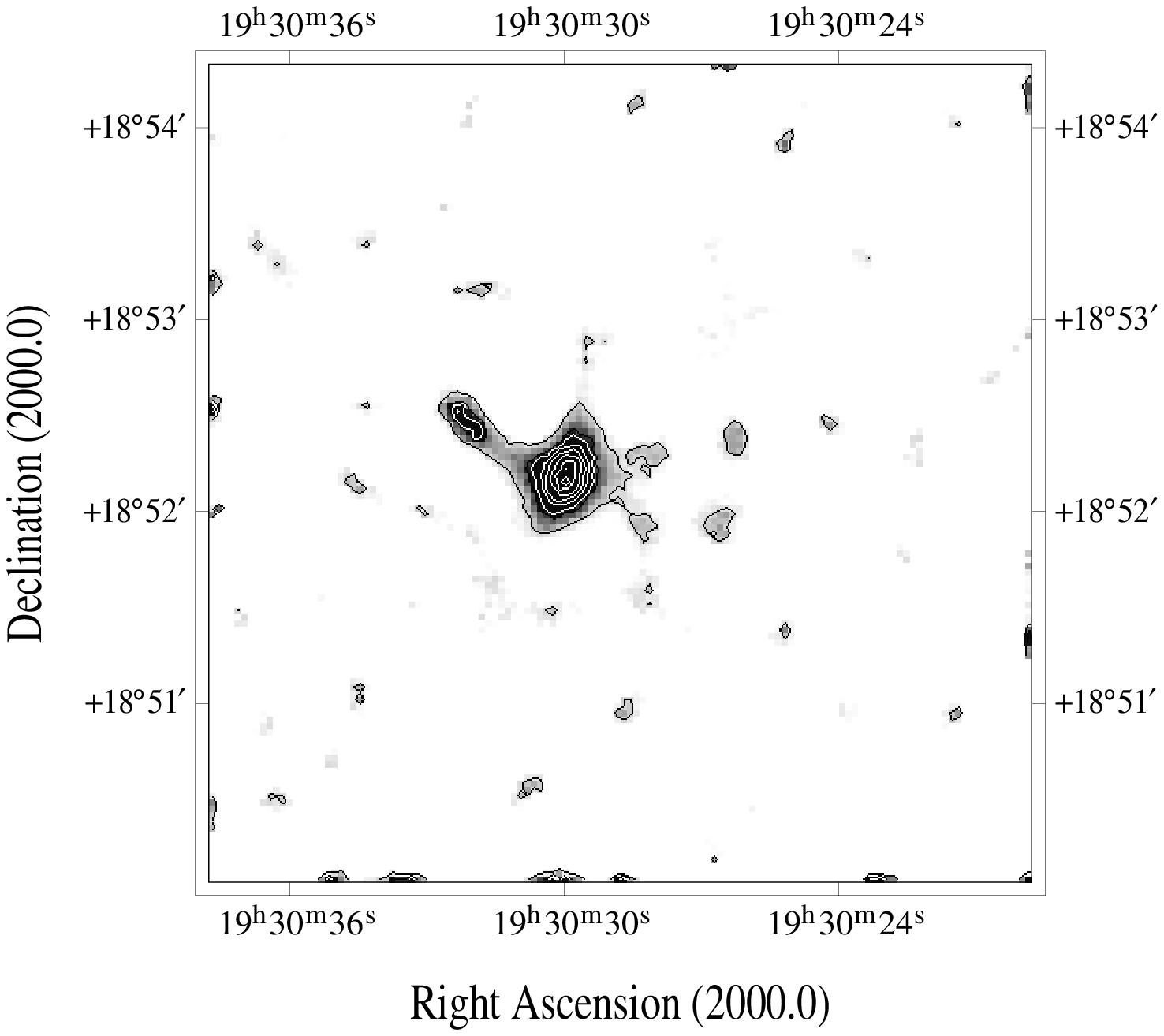,width=8.8truecm,angle=0,%
   bbllx=65pt,bblly=180pt,bburx=458pt,bbury=544pt,clip=}
\caption{Restorated 0.1-2.5 keV X-ray intensity map of G54.1+0.3.
 North is up and east is left. The contour intervals are linear with a step size
of 4$\times$10$^{-5}$ cnt s$^{-1}$ per 2$\arcsec\times$2$\arcsec$ pixel.
 The lowest
contour corresponds to a brightness level of 
2$\times$10$^{-5}$ cnt s$^{-1}$ per pixel.}
\label{picture}
\vspace{0cm}
\end{figure}

In order to exam the reliability of the restored image, we have performed a
Monte-Carlo simulation. Figure 6(a) displays an object similar to 
G54.1+0.3 in figure 4 in shape and flux. Figure 6(b) is the simulated $ROSAT$ 
PSPC observational 
result with the same observing time and background level as the 
real observation to G54.1+0.3 , and figure 6(c) 
is the smoothed image from 6(b). Figure 6(d) is the restorated image of
6(b). The simulation shows that the high resolution X-ray image of 
G54.1+0.3 we obtained is reliable. 
\begin{figure}[thb]
 \psfig{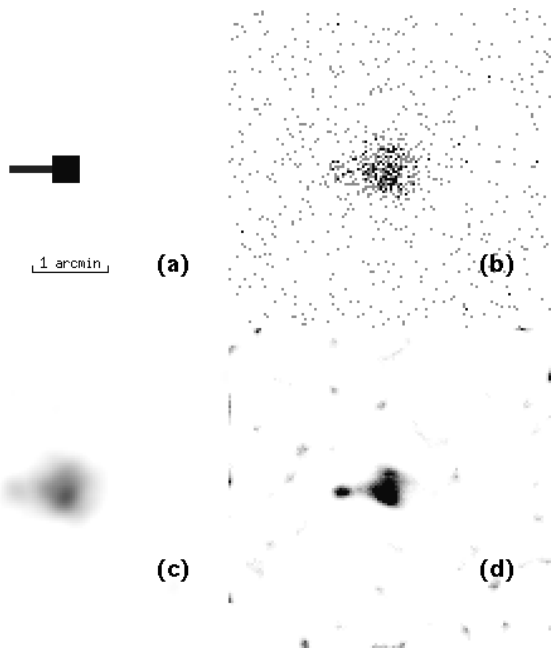}
\caption{Simulation of the $ROSAT$ PSPC observation and image restoration
to an object similar to G54.1+0.3. (a) Object which has a similar shape and the
same flux with G54.1+0.3. (b) Simulated $ROSAT$ PSPC observation results 
with the same intergrated time and background level. (c) A smoothed image 
from (b) and an 10$\arcsec$ FWHM Gaussian filter. (d) The restorated image.
Each of the four images has the similar angular size with figures 1 and 4.
Note that noisy features have not been displayed in (c) because of a high
cut level.}
\label{picture}
\vspace{0cm}
\end{figure}

\section{Discussions}
\subsection{Distance and X-ray luminosity of G54.1+0.3}
Velusamy \& Becker (1988) suggested that G54.1+0.3 may have a distance
of about 3.2 kpc, if its progenitor is in the star-forming region
G53.9+0.3. The galactic HI column density in this direction is about
14.5$\times$10$^{21}$ cm$^{-2}$ (Dickey $\&$ Lockman 1990).
 The best fit column density we get from 
$ROSAT$ PSPC observation is a little lower  and the best fit column 
density of the $ASCA$ SIS 
observation is a little higher than that value. These  
column densities imply a distance comparable with the radius of
the galaxy, similar to the result of $EINSTEIN$ IPC (Seward 1989), 
3.2 kpc might be then too close and 10 kpc should be a reasonable 
estimation.  
The X-ray luminosity in 0.1-2.4 keV 
is $L_X$ = 3.2$\times$10$^{33} d_{10}^{2}$ erg s$^{-1}$, 
where $d_{10}$ is the distance to G54.1+0.3 in unit of 10 kpc.
If the distance does not deviate from 10 kpc very much, its X-ray
luminosity is three or four orders' lower than
Crab Nebula (Helfand \& Becker 1987), lies in the lower end of Crab-like 
SNRs, similar to that of 
SNR 3C58 (Helfand \& Becker 1987; Helfand et al. 1995). 

The radio luminosity
of G54.1+0.3 is about 5$\times$10$^{33}$ $d_{10}^{2}$ erg s$^{-1}$,  
 derived from the radio observations of 
Velusamy \& Becker (1988). The ratio $L_x/L_r$ = 0.6, also similar to
that of 3C58 (Helfand \& Becker 1987).  

Seward \& Wang (1988) found that a relation between 
the X-ray luminosity ($L_X$) of a 
plerionic SNR and the spin-down luminosity ($\dot{E}$) of the central 
pulsar. Using that relation we can derive 
$\dot{E}\sim$8$\times$10$^{35}$ erg s$^{-1}$  
for the central pulsar in G54.1+0.3.

\subsection{Electron energy distribution} 
The $ROSAT$ PSPC observation of G54.1+0.3 shows that the X-ray flux at
1 keV  is about 6.4$\times$10$^{-4}$ mJy. As the radio flux at 1.4 GHz
is 478 mJy, the flux index between radio and X-ray is about -0.7,
a little flatter than the $ASCA$ SIS obtained X-ray energy index (-0.9)
and much steeper than that of radio spectral index (-0.13) 
(Velusamy $\&$ Becker 1998), indicating that the spectrum contains a 
break between radio and X-ray. Comparing the radio to X-ray flux index,
 radio flux index and the X-ray flux index, we find that the break
 is around 10$^{11}$ Hz. 
 
  If the relativistic electrons
have a power law energy distribution $n_e$ = $E^{\gamma}$, the spectral index 
$\alpha$ is $\frac{\gamma+1}{2}$. The spectral break in 
the spectral means a similar break in the electron energy distribution.
The critical radiation frequency of a relativestic electron with 
energy $E$ in a magnetic field with strength $B$ is 
$\nu_c = 16.1BE^{2}sin\psi$ MHz, where $B$ is in $\mu$G, $E$ is in
GeV and $\psi$ the incident angle of electron (Lang 1998). 
The electrons whose maximum radiations are at 
1 GHz have typical energies of 9.4$B^{-0.5}$ GeV with $B$ the magnetic 
field strength in $\mu$G, assuming that the incident angle is
45$\degr$. Similarly the X-ray (around 1 keV) emitting relativestic 
electrons will have typical energies of 94$B^{-0.5}$ TeV. If the
magnetic field is about 10$\mu$G, the above estimations show that
the electron energy distribution is $\sim$$E^{-1.3}$ around 30 GeV
and $\sim$$E^{-2.8}$ around 300 TeV. An index break exists between
30 GeV and 300 TeV, probably around 300 GeV. 

The life time of a relativestic electron can be represented by $t_{1/2}$, the
time of the electron loses half of its initial energy $E_0$, 
$t_{1/2}=\frac{8.35\times10^{9}}{(B sin\psi)^{2}E_0}$ years, 
where $B$ is in $\mu$G,
$E_0$ is in GeV and $\psi$ the incident angle of the electron (Lang 1998). 
The lifetimes for the 30 GeV, 300 GeV and 300 TeV photons in 10 $\mu$G
magnetic field are about 5.6$\times$10$^{6}$, 5.6$\times$10$^{5}$, 5.6$\times$10$^{2}$
years, respectively. These three typical lifetimes will be used
in the discussions of the origin of the electron energy distribution 
index break in the next paragraph.

If the electrons from the center pulsar have a continuous power law energy
distribution initially, the observed break should be due to 
the short lifetime of the
high energy electrons. Because the low energy 
electrons which radiate radio emission have a long lifetime, their energy 
distribution represents the initial electron energy distribution well.
The initial energy flux ratio $\frac{f_{1.4GHz}}{f_{1keV}}$ is then 11.75. 
The currently 
observed energy flux ratio $\frac{f_{1.4GHz}}{f_{1keV}}$ is 7.8$\times10^{5}$,
indicates that the age of G54.1+0.3 would be at least 
$\frac{7.8\times 10^{5}}{11.75}$$\times$5.6$\times$10$^{2}$=3.7$\times$10$^{7}$
years. This large age value shows that the observed electron energy 
distribution break is quite probably an intrinsic property of the
electrons from the central pulsar.

\subsection{X-ray jet}
For the first time an well resolved X-ray image of G54.1+0.3 has been obtained. 
It shows a JLF
 pointing to the northeast. The simulation shows that such a 
structure can be clearly resolved by $ROSAT$ PSPC with the aid of 
an image restoration technique. The simulation also shows that this feature
can not be attributed as the fluctuations of the bright source, it is an
intrinsic structure of the object. 

We have studied the possibility that the JLF is indeed a separate object
lies in a similar direction with G54.1+0.3. We find that there is no
identified object in the 30$\arcsec$ vicinity of the JLF except 
G54.1+0.3. The optical plate obtained by Palomar Observatory Sky Survey
and electronically reproduced by Skyview of NASA/GSFC shows no source
in the JLF region too. The JLF shown in figure 5 shows some enhancements
in the head. But it might be a false phenomenon caused by the low quality
of the original data and the restoration process, as can be found in the  
simulation, although some similar structures exist in the 4.8 GHz radio map. 
More simulations show that the length of the JLF is quite reliable, the
width of the JLF might have an uncertainty up to $\sim$ 50$\%$.

We have compared figure 4 with the 4.8 GHz VLA map obtained by 
Velusamy \& Becker (1988) in details. The brightest point of the extended 
X-ray source locates at R.A.(2000)=19:30:30.0, 
DEC(2000)=18:52:07,
which coincides with the brightest region of the radio source. 
The head of the JLF
has a coordinate of R.A.(2000)=19:30:32.2,
DEC(2000)=18:52:31,
which also coincides with
the northeastern enhancement in the radio map. 
The nice position coincidence of the X-ray
and radio sources strongly favor their same origin. However, the X-ray 
source has a smaller extent than the radio source and no significant X-ray  
emission has been detected along the northward feature, which was suggested
to be the most probable radio JLF by Velusamy \& Becker (1988). It might 
be due to
the intrinsic deficiency or the limited sensitivity of the present observation.
 
There are two possible ways to explain the origin of the X-ray JLF. 
One is that it is 
a fragment produced in the supernova explosion, like the fragments detected 
around the Vela SNR, especially its `bullet'-like fragment A. (Aschenbach et al. 1995; 
Strom et al. 1995). However, significant radio emission has only been detected
around the head of the fragments, implies that most of the
relativistic electrons are in the leading edge of the fragments, close to the 
shock front (Strom et al. 1995). But in the case of 
G54.1+0.3 the radio emission has a similar distribution with the X-ray JLF,
indicating a similar distribution of relativistic electrons with the 
X-ray brightness. It makes the fragment origin of the X-ray 
JLF implausible. The second is that the X-ray JLF is due to the relativistic
electrons produced by the central pulsar, like X-ray jets detected in 
PSR 1929+10 (Wang et al. 1993), Crab 
SNR (Hester et al. 1995), Vela pulsar (Markwardt \& \"Ogelman 1995), 
 SNR MSH 15-52 (Tamura et al. 1996), SNR CTB80 (Wang \& Seward 1984;
Safi-Harb et al. 1995) in the galaxy and SNR N157B in the 
Large Magellanic Cloud (Wang \& Gotthelf 1998). 
The coincidence of radio and X-ray emission in the case of Vela 
pulsar jet (Frail et al. 1997) and that 
of SNR N157B (Wang \& Gotthelf 1998) strongly support this scenario. 
We conclude that the JLF we discoveried is quite probably an X-ray jet connected
with the pulsar in G54.1+0.3. 

The X-ray emission of Vela pulsar jet can be 
fitted with both power law and thermal plasma model (Markwardt \& \"Ogelman 1995),
and the X-ray pulsar jet in MSH15-52 appears to be nonthermal. It is difficult
to get the spectral properties of the X-ray jet in G54.1+0.3 with the present
data.  We assume that it share the same power law model with the
whole remnant, and is due to the synchrotron radiation of relativistic
electrons from the pulsar. The X-ray luminosity of the jet in 0.1-2.4 
keV is then  about
5.1$\times$10$^{32}$$d_{10}^2$ erg s$^{-1}$. 

From the radio map of Velusamy $\&$ Becker (1988)
we estimate that the flux of the jet at 4.8 GHz is about 40 mJy. 
Its X-ray flux at 1 keV is about 9.4$\times10^{-5}$ mJy.
The two fluxes give a spectral index from radio to X-ray of about -0.73,
quite similar to that of the whole remnant. As no significant radio spectral
variation across the source has been detected (Velusamy $\&$ Becker 1988),
the jet electrons have a break with the energy distribution too, similar to
the whole remnant.

The distance of the jet head to the nebula center is about 40${\arcsec}$.
It corresponds to 2 pc if the SNR is 10 kpc away. Reccent distance measurements
to Vela SNR obtained a distance of 250$\pm30$ pc (Cha et al. 1999). If so 
the Vela pulsar jet is about 3 pc 
long (Cha et al. 1999, Markwardt \& \"Ogelman 1995). The lengths of the two 
jets are quite similar.

\section{Summary}
$ROSAT$ PSPC and $ASCA$ observations of G54.1+0.3 imply a large distance 
comparable with the galactic radius. Its X-ray spectrum is of nonthermal
 origin. 
The comparison of the radio and X-ray emissions shows that the energy 
distribution of the relativistic electrons has a
break around 300 GeV. This break is quite probably an intrinsic property
of the relativistic electrons from the central pulsar instead of due to
the energy lose in the synchrotron radiation process, if 
G54.1+0.3 is not as old as 3.7$\times$10$^{7}$ years.  

 A high
spatially resolved image shows an X-ray jet pointing to the northeast,
similar to the radio structures. Its nonthermal spectrum and the 
existence of X-ray jet confirm the formal identification of G54.1+0.3 as
a Crab-like SNR, though no pulsation has been found in the X-ray observation. 
Future deep X-ray observations with high spatial resolution
and spectral resolution telescopes such as Chandra and XMM are invaluable to
find out the spectral and spatial structure of the remnant as well as the
X-ray jet.
 
\begin{acknowledgements}
F.J. Lu is supported by 
the exchange program between 
Max-Planck Society and Chinese Academy of Sciences. He thanks
Professor J. Tr\"umper for hospitality. The authors  
thank Drs S.D. Mao and Q.D. Wang for helpful discussions. 
This research is partially supported by the National Natural Science Foundation 
of China and the Special Funds for Major State Basic Research Projects. It
has
made use of the SIMBAD  database, operated at CDS, Strasbourg, France
and the Digitized Sky Survey operated by Skyview of NASA/GSFC.

\end{acknowledgements}


\begin{thebibliography}{}
   \bibitem[1981]{an81}Angerhofer P.E., Strom R.G., Velusamy T., Kundu M.R.,
   1981, A\&A 94, 313
   
   \bibitem[1995]{as95}Aschenbach B., Egger R., Tr\"umper J., 1995, Nat 373, 587
  
   \bibitem[1999]{ch99}Cha A.N., Sembach K.R., Danks A.C., 1999, ApJ 515, L25
   
   
   \bibitem[1990]{di90}Dickey J.M., Lockmann F.J., 1990, ARA\&A 28, 215
   
   \bibitem[1997]{fr97}Frail D.A., Bietenholz M.F., Markwardt C.B., 
   \"Ogelman H., 1997, ApJ 475, 224   

   \bibitem[1987]{he87}Helfand D.J., Becker R.H., 1987, ApJ 314, 203
   
   \bibitem[1995]{he95}Helfand D.J., Becker R.H., White R.L., 1995, ApJ 453, 741

   \bibitem[1995]{he95}Hester J.J. et al., 1995, ApJ 448, 240
 
   \bibitem[1998]{la98}Lang K.R., 1998, Astrophysical Formulae (Springer) 

   \bibitem[1994]{li94}Li T.P., Wu M., 1994, Ap\&SS 215, 213
   
   \bibitem[1996]{lu96}Lu F.J., Li T.P., Sun X.J., Wu M., Page C.G., 1996,
   A\&AS 115, 395
   
   \bibitem[1974]{lu74}Lucy L., 1974, AJ 79, 745
   
   \bibitem[1995]{ma95}Markwardt C.B., \"Ogelman H., 1995, Nat 375, 40
   
   \bibitem[1977]{ra77}Raymond J.C., Smith B.W., 1977, ApJS 35, 419
   
   \bibitem[1985]{re85}Reich W., F\"urst E., Altenhoff W.J., Reich P., 
   Junkes N., 1985, A\&A 151, L10
   
   \bibitem[1972]{ri72}Richardson B.M., 1972 J. Opt. SOc. Am. 62, 55

   \bibitem[1985]{ro85}Roger R.S., Milne D.K., Kesteven M.J., Haynes R.F.,
   Wellington,K.J., 1985, Nat 316, 44
     
   \bibitem[1995]{sa95}Safi-Harb S., \"Ogelman H, 
    Finley J.P., 1995, ApJ 439, 722
 
   \bibitem[1989]{se89}Seward F.D., 1989, AJ 97, 481
   
   \bibitem[1988]{se88}Seward F.D., Wang Z.R., 1988, ApJ 332, 199
   
   \bibitem[1995]{st95}Strom R., Johnston H.M., Verbunt F., Aschenbach B., 1995,
   Nat 373, 590
  
   \bibitem[1996]{ta96}Tamura K., Kawai N., Yoshida A., Brinkmann W., 1996,
   PASJ 48, L33
  
   \bibitem[1994]{ta94}Tanaka Y., Inoue H., Holt S.S., 1994, PASJ 46, L37
  
   \bibitem[1984]{ve84}Velusamy T., 1984, Nat 308, 15
   
   \bibitem[1988]{ve88}Velusamy T., Becker R.H., 1988, AJ 95, 1162
   
   \bibitem[1998]{wa98}Wang Q.D., Gotthelf E.V., 1998, ApJ 494, 623
   
   \bibitem[1993]{wa93}Wang Q.D., Li Z.Y., Begelman M.C., 1993, Nat 364, 127
   
   \bibitem[1984]{wa84}Wang Z.R., Seward F.D., 1984, ApJ 285, 607

   \bibitem[1998]{zh98}Zhang S., Li T.P., Wu M., 1998, A\&A 340, 62
     
   \bibitem[1998]{zi98}Zimmermann U., Boese G.,  Becker W., Belloni T.,
D\"obereiner S., Izzo C., Kahabka P., Schwentker O. 1998, 
$EXSAS$ $user's$ $guide$, MPE Report

\end{thebibliography}
\end{document}